\documentstyle[12pt,epsf]{article}

\begin{document}

\title{Spreading of Damage in the SK Spherical Spin Glass}
\author{Daniel A. Stariolo\\[0.5em] 
{\small Dipartimento di Fisica, Universit\`a di Roma I} 
{\small {\em ``La Sapienza''}}\\
{\small and Istituto Nazionale di Fisica Nucleare, Sezione di Roma I}\\
{\small Piazzale Aldo Moro 2, 00185 Rome, Italy}\\[0.3em]}
\maketitle

\begin{abstract}
By considering the Langevin dynamics of
the SK spin glass with a spherical constraint we calculate the asymptotic
distance between two real replicas that evolve with the same thermal noise
from different initial conditions. Despite the simplicity of the model its
dynamics is known to be non trivial and presents aging phenomena. We found
that the asympotic distance between two replicas is maximal for a whole family
of initial conditions that are arbitrarily close at $t=0$. 
The asymptotic distance presents a dynamical transition at a temperature
equal to the static phase transition one.

\end{abstract}
\thispagestyle{empty}
\newpage

\section{\bf Introduction}

The sensibility of the dynamics of a spins system with respect to the initial
conditions has been extensively studied in the last years under the name of
{\it spreading of damage}. This name was proposed in connection with
biological systems where an important issue is the propagation of a little
mutation or damage and its stability in an otherwise ``normal'' system 
(Kauffman 1969). In a spin system, information about the structure of phase
space can be gained by observing the temporal evolution of the system from
different initial conditions. Suppose we have two replicas of a system,
 ${\bf
\sigma}$ and ${\bf \tau}$, and let them evolve in exactly the same external
conditions (thermal noise, fields).
The question then is: Are the long time dynamics of the two replicas 
sensible to small differences in initial conditions? This has been extensively
studied in cellular automata (Martins, Verona de Resende, Tsallis and de
Magalh\~aes 1991, Hinrichsen, Weitz and Domany 1996), neural networks (Tamarit
and Curado 1994),
 ferromagnetic Ising models (Stanley, Stauffer, Kertesz and Herrmann 1987,
 Derrida and Weisbuch 1987) and spin glasses (Derrida 1989). The
quantities of interest are the Hamming distance at time {\it t} between the
two replicas:
\begin{equation}
D(t)=\frac{1}{4N}\sum_{i=1}^N (\sigma_i(t) - \tau_i(t))^2
\end{equation}
and the probability $P(t)$ that the two replicas remain different after a
time {\it t}.
Different dynamical regimes or dynamical phase transitions can be seen by
the change in the behaviour of $D(t)$ for $t\rightarrow\infty$. For example
in the heat bath dynamics of a ferromagnet the distance goes to zero in the
paramagnetic phase and it becomes different from zero below $T_c$ if the
initial configurations of the replicas correspond to different magnetization
states. A little damage always heals in a heat bath ferromagnet 
(Vojta 1996 a,b). This result has a trivial interpretation in terms of the 
structure of phase space. 

In complex systems like spin glasses the phenomenology observed, mainly
through computer simulations, is much more interesting (Derrida 1989). In
general, three asymptotic regimes have been observed. A high temperature
regime where the distance vanishes signalling a simple phase space with only
one state down to a temperature $T_1$. This $T_1$ happens to coincide or 
be slightly smaller than the critical temperature $T_c$ of the pure 
ferromagnetic
system. For $T_2 < T < T_1$ the distance goes to a nonzero value that is
independent of the initial conditions on the replicas. Finally for $T < T_2$
the distance goes to a nonzero value dependent on the initial conditions.
This second temperature is near the spin glass transition temperature. 
Clearly this behaviour as $T$ is lowered reflects the increasing complexity
of the phase space encountered by the replicas in their dynamics.

Extensive computer simulations done 
 in spin glasses in dimensions 2, 3, 4 and 6 
and mean field (Campbell and de Arcangelis 1991),
and in spin glasses with asymmetric interactions 
(de Almeida, Bernardi and Campbell 1995), suggest that there is no dependence
at all on the initial conditions, but this is probably because the simulations
 have not reached a sufficiently low temperature remaining at $T \geq T_2$. 

Another important fact that may constrain the final output is the choice of 
the effective temperature of the initial conditions. Choosing random initial
conditions that correspond to infinite temperature, the system will go 
through a rapid quench to the final temperature $T$. In this way it will
select at random one of the accesible states at that temperature, probably
remaining in those of higher energy for extremely long times. Other
possibility is to initialize the replicas in equilibrium configurations at
temperatures $T$ and $T'$.

Despite the efforts done through extensive computer simulations, there are
few analytic appoaches to the spreading of damage. Golinelly and Derrida
(Golinelli and Derrida 1988) calculated the survival probability $P(t)$ for a 
mean field
ferromagnet and Derrida (Derrida 1987) obtained the solutions for $D(t)$ in
spin glasses with strongly diluted asymmetric interactions. Recently Vojta
obtained interesting results for ferromagnets from a master equation
approach in an effective field approximation (Vojta 1996 a,b).
Trying to make contact with thermodynamic quantities,
Coniglio, de Arcangelis, Herrmann and Jan (1989) have found 
relations between equilibrium 
spatial correlation functions and the (equilibrium) Hamming distance. Our
aim in this work is to exploit the relation between the Hamming distance and
dynamical, i.e. time dependent, correlation functions, as a way of extracting
information of the precise dynamics of the spreading of damage process in
complex systems. We have found analytic expressions for the distance $D(t)$
in a simple spin glass model, namely the SK model with a spherical 
constraint. This model is known to have a thermodynamic phase transition to
a low temperature phase with replica symmetry (Kosterlitz, Thouless and Jones
1976). 
 Thus it cannot be considered a
``true'' spin glass and its phase space structure can be analyzed with great 
detail, at least for $T\rightarrow 0$ (Kurchan and Laloux 1996).
Despite its relative simplicity,
 a detailed study of the off equilibruim dynamics of the model
(Cugliandolo and Dean 1995) revealed interesting properties like aging. We have studied
 the spreading of damage in this model finding some interesting peculiarities
of its phase space
as the exponential separation of some initially arbitrary near 
trajectories at zero temperature. On general grounds we have found a strong 
dependence  of the final distance on the initial conditions and the absence
of the intermediate regime that is consequence of the simple
structure of phase space.

\section{\bf The Model}

The spherical SK model is defined by the Hamiltonian:
\begin{equation}
H=-\frac{1}{2}\sum_{i\ne j} J_{ij}\sigma_i \sigma_j
\end{equation}
where ${\sigma_i, i=1\ldots N}$ are real valued spin variables that 
satisfy the
constraint $\sum_{i=1}^N \sigma_i^2 =N$. The couplings between the spins are
given by the quenched random variables $J_{ij}$.

A possible dynamics for the model is defined by a Langevin equation which
in the basis of the eigenvectors of the interaction matrix reads:
\begin{equation}
\frac{\partial \sigma_{\mu}(t)}{\partial t}=(\mu-z(t))\sigma_{\mu}(t)+
\xi_{\mu}(t) \label{dynamics}
\end{equation}
where $\mu$ is the eigenvalue associated with the $\mu^{th}$ eigenvector, 
$z(t)$ is
a Lagrange multiplier enforcing the spherical constraint and $\xi_{\mu}(t)$
is a thermal white noise with zero mean and correlations:
\begin{equation}
\langle \xi_{\mu}(t)\xi_{\nu}(t')\rangle=
2T\delta_{\mu\nu}\delta(t-t')
\end{equation}
$T$ is the temperature of the heat bath and $\langle\cdots\rangle$ means an
average over the thermal noise.

The general solution of Eq.(\ref{dynamics}) is (Cugliandolo et al. 1995):
\begin{eqnarray}
\sigma_{\mu}(t)& =& \sigma_{\mu}(t_0)e^{\mu(t-t_0)}\exp{\left\{
-\int_{t_0}^t d\tau\,z(\tau)\right\}}+\\ \nonumber
               &  & \int_{t_0}^t dt'' e^{\mu(t-t'')}\exp{\left\{
-\int_{t''}^t d\tau' \,z(\tau')\right\}} \xi_{\mu}(t'') \label{sigma}
\end{eqnarray}

In this work we are interested in the time evolution of two real replicas
of the system, $\bf \sigma$ and $\bf \tau$, that evolve with the same thermal
noise. We then define the overlap between
them at time $t$:
\begin{equation}
Q(t)=\frac{1}{N} \left[ \sum_i \langle \sigma_i(t) \tau_i(t) 
      \rangle \right]_J
\end{equation}
where $[\cdots]_J$ means average over the disorder variables $J_{ij}$.
In the basis of the eigenvectors of {\it J} it can be expressed as
\begin{eqnarray}
Q(t) & \equiv & \left[ \langle \sigma_{\mu}(t) \tau_{\mu}(t) \rangle
                    \right]_J        \nonumber \\
     & =      & \int d\mu \rho(\mu) \langle \sigma_{\mu}(t)\tau_{\mu}(t)
                \rangle
\end{eqnarray}
where $\rho(\mu)$ is the density of eigenvalues of the interaction matrix {\it J}.
A common choice is an ensemble of matrices with 
independent gaussian elements with
zero mean and variance proportional to $1/N$. In the limit of large $N$ the 
eigenvalues
of this ensemble are distributed according to the Wigner semicircle law:
\begin{equation}
\rho(\mu)=\frac{1}{2\pi}\sqrt{4-\mu^2} \hspace{2cm} \mu \in [-2,2].
\end{equation}
Using the result for $\sigma_{\mu}(t)$ (Eq.(\ref{sigma})) (setting $t_0=0$) 
 one obtains an expression for the overlap
\begin{equation}
Q(t)=\frac{1}{\sqrt{\Gamma_{\sigma}(t)\Gamma_{\tau}(t)}}\left[
     \sigma_{\mu}(0)\tau_{\mu}(0)e^{2\mu t} + 2T\int_0^t dt' 
     \sqrt{\Gamma_{\sigma}(t')\Gamma_{\tau}(t')} e^{2\mu (t-t')} \right]_J
	\label{overlap}
\end{equation}
where $\Gamma(t)\equiv \exp{(2\int_0^t dt' z(t'))}$. Using the spherical constraint
$\Gamma(t)$ can be computed as the solution of a Volterra integral equation of the second kind (Cugliandolo et al. 1995):
\begin{equation}
\Gamma(t)=\left[ (\sigma_{\mu}(0))^2 e^{2\mu t} + 2T \int_0^t dt' \Gamma(t')
          e^{2\mu (t-t')} \right]_J       \label{gama}
\end{equation}
The spherical constraint implies $\Gamma(0)=1$.

By solving the closed equations (\ref{overlap}) and (\ref{gama}) for different
sets of initial conditions and using the spherical constraint we can finally 
calculate the distance between the two replicas at time t as:
\begin{equation}
D(t)=\frac{1-Q(t)}{2}  \label{distance}
\end{equation}
which is a direct generalization of the Hamming distance commonly used in binary
systems.

\section{\bf Results}

We have studied the asymptotic behaviour of the distance $D(t)$ for three 
classes of initial conditions, namely:
\begin{enumerate}
\item Opposite initial conditions: $\sigma_{\mu}(0)=-\tau_{\mu}(0)$.
\item Independent initial conditions: $\left[ \sigma_{\mu}(0)\tau_{\mu}(0)
      \right]_J =0$.
\item Arbitrarily correlated initial conditions. This class includes as particular
case initial conditions that differ by an arbitrarily small fraction of spins
as in the original defintion of damage.
\end{enumerate}
In the first two classes we can even distinguish the initial conditions in two
additional sub-classes:
\begin{itemize}
\item Uniform initial conditions: $\sigma_{\mu}(0)=1, \forall \mu$. This has the
same projection onto each eigenvector of {\it J} and
corresponds to a random configuration in the original basis.
\item Staggered initial conditions: $\sigma_{\mu}(0)=\sqrt{N}\delta_{a\mu}$.
A condition of this type has a finite projection onto only one eigenvector of {\it
J}.
\end{itemize}
We now proceed to present the results for each case.

\subsection{Opposite Initial Conditions}
First of all we can see immediatly that if $\sigma_{\mu}(0)=\tau_{\mu}(0)$ then
$\Gamma_{\sigma}=\Gamma_{\tau}$ and the distance is zero for all t, as it must.
Now if $\sigma_{\mu}(0)=-\tau_{\mu}(0)$ we still have $\Gamma_{\sigma}(t)=
\Gamma_{\tau}(t)\equiv \Gamma(t)$, but from Eqs.(\ref{overlap}) and (\ref{gama})
\begin{equation}
Q(t)=1-\frac{2}{\Gamma(t)}\left[ (\sigma_{\mu}(0))^2 e^{2\mu t} \right]_J
\end{equation}

\subsubsection{Uniform initial condition}
In this case we have :
\begin{equation}
\left[ (\sigma_{\mu}(0))^2 e^{2\mu t}\right]_J = \frac{I_1(4t)}{2t}
\end{equation}
and
\begin{equation}
\Gamma(t)=\frac{1}{T}\sum_{k=0}^{\infty}k\frac{I_k(4t)}{2t} T^k \hspace{1cm}
                                                                T < 1.
\end{equation}
where $I_k(x)$ is the modified Bessel function of order $k$. 
For large arguments
$I_k(x) \approx e^x/\sqrt{2\pi x}$ and $Q(t)$ tends to the asymptotic value
$\lim_{t\rightarrow\infty}Q(t)\equiv Q_{\infty}= 1-2(1-T)^2$. Consequently the 
asymptotic distance between the two initially opposite replicas will be:
\begin{equation}
D_{\infty} = (1-T)^2 = q_{EA}^2
\end{equation}
where $q_{EA}$ is the Edwards-Anderson order parameter that equals $1-T$ in the
spherical spin glass (Kosterlitz et al. 1976). At $T=0$ the distance is frozen and
$D(t)=D(0)=1$ for all times. This can be interpreted within a geometric 
description of phase space (Kurchan et al. 1996): the energy has two absolute minima parallel and antiparallel in the 
direction of the maximum eigenvalue of $J_{ij}$, $\mu=2$. Phase space is divided
in two equivalent halves that are geometrically identical and form the basins
of attraction of the two minima. Consequently two opposite uniform initial 
conditions will evolve to the two distinct minima encoutering in the evolution
the same landscape and so preserving the distance along the way towards the
minima, which in turn dominate the asymptotic dynamics. 

As $T$ grows the thermal fluctuations permit the replicas
to come closer to each other up to a value equal to the square of the self-overlap
and finally merge when $T$ goes to one signalling the transition to a high temperature
phase characterized by $D=0$, that means the system has a unique pure state.

\subsubsection{Staggered initial condition}

In this case $[(\sigma_{\mu}(0))^2 \exp{(2\mu t)}]_J=\exp{(2at)}$. $\Gamma(t)$ can't
be expressed in a simple closed form as in the previous case so
we only quote its asymptotic form that depends on the value of $a$. If $a=2$ (the
initial condition has a nonvanishing projection onto the maximum eigenvalue) then
$\Gamma(t)\approx \exp{(4t)}/(1-T)$ and
\begin{equation}
D_{\infty} = 1-T = q_{EA}
\end{equation}
The distance decreases linearly with $T$. 
Comparing with the previous case, even
though in both cases the initial conditions are opposite, when the systems 
start
from a random configuration the final distance is smaller than when they start
aligned in the direction of the maximum eigenvalue. As was noted by 
Kurchan et al. (1995), at finite temperature the component in the direction of the
maximum eigenvalue has a finite probability of changing sign at any time unless
the initial condition is deep in a basin, this fact may be responsible for the
different asymptotic distance between uniform and staggered ($a=2$) initial
conditions.

If $a < 2, \Gamma(t)\approx \frac{T}{(1-T)^2}\frac{1}{2-a}\frac{1}{\sqrt{4\pi}}
 \frac{\exp{(4t)}}{(2t)^{3/2}}$ and the overlap asymptotically behaves as
\begin{equation}
Q(t)\approx 1-4\sqrt{\pi}\frac{(1-T)^2}{T} (2-a)(2t)^{3/2} e^{(2a-4)t},
\end{equation}
so it relaxes exponentially to 1. This means that in the infinite time limit 
the distance between the replicas goes to zero for all $T > 0$. This again can
be attributed to the fact that the initial conditions are not deep in the 
basins of the minima and in this case
, although opposite, have in fact zero initial projection on them. 

If $T=0 $ we can see
from (\ref{overlap}), (\ref{gama}) and (\ref{distance}) that $D(t)=1$ at all 
times (the distance is frozen). Staggered initial conditions are fixed points
of the dynamics at zero temperature and consequently the distance between them
remains maximal.

\subsection{Independent initial conditions}

Independent initial conditions satisfy $\int d\mu \rho(\mu) \sigma_{\mu}(0)
\tau_{\mu}(0) = 0$. We have studied two cases, one with staggered initial 
conditions ($a < 2$)in both replicas and other with uniform projection on a subset of
the eigenvectors of {\it J} in such a way as to preserve independence between the
initial configurations. In both cases $ \left[ \sigma_{\mu}(0)\tau_{\mu}(0)
\exp{(2\mu t)}\right]_J = 0$ and the equation for the
overlap reads:
\begin{equation}
Q(t)=\frac{1}{\sqrt{\Gamma_{\sigma}(t)\Gamma_{\tau}(t)}}
      2T\int_0^t dt' 
     \sqrt{\Gamma_{\sigma}(t')\Gamma_{\tau}(t')} \left[ e^{2\mu (t-t')} \right]_J
	\label{overindep}
\end{equation}

\subsubsection{Staggered initial conditions}

Let's consider staggered initial conditons in both replicas:
\begin{eqnarray}
\sigma_{\mu}(0) & = & \sqrt{N}\delta_{a\mu} \nonumber \\
\tau_{\mu}(0)   & = & \sqrt{N}\delta_{b\mu} \hspace{2cm} a\neq b \neq 2.
\end{eqnarray}
In this case $\left[ (\sigma_{\mu}(0))^2 \exp{(2\mu t)}\right]_J = \exp{(2at)}$ and
$\left[ (\tau_{\mu}(0))^2 \exp{(2\mu t)}\right]_J = \exp{(2bt)}$. 
We have not been able to solve the integral for the overlap Eq.(\ref{overindep})
analytically but we have done the
integration numerically (Linz 1985). The result for the asymptotic distance as a
funtion of temperature is shown in Fig.(1) for two arbitrary initial
conditions in different basins.


Again at $T=0$ the dynamics is frozen and $D(t)=D(0)=0.5$. Note that in the case of
opposite initial conditions the initial overlap was $Q(0)=-1$ and now it is 
$Q(0)=0$. At an arbitrarily small but finite temperature the final distance falls
down to small values and already for $T \ge 0.05$ we have $D_{\infty} \le 0.1$. This means
that although the initial configurations are uncorrelated they finally become very
near each other, although they do not completely merge as was the case with 
opposite initial conditions.

\subsubsection{Piecewise uniform initial conditions}

We choose here initial conditions of the form:
\[
\sigma_{\mu}(0)=\left\{ 
\begin{array}{r@{\quad if \quad}l}
\kappa &   \mu < c \\
0      &   \mu \ge c  
\end{array} \right.
\]
and
\begin{equation}
\tau_{\mu}(0)=\left\{ 
\begin{array}{r@{\quad if \quad}l}
0 &  \mu < c \\
\lambda  &  \mu \ge c  
\end{array} \right.
\end{equation}
The values of $\kappa$ and $\lambda$ depend on $c$ through the spherical constraint:
$\int_{-2}^c d\mu \rho(\mu) \kappa^2 =1$ and $\int_c^2 d\mu \rho(\mu) \lambda^2 =1$.
For simplicity we have chosen $c=0$ so that $\kappa^2 = \lambda^2 = 2$. For the
Gamma functions we obtained the following equations:
\begin{eqnarray}
\Gamma_{\sigma}(t) & = & e^{-2t}\frac{I_1(2t)}{t}+ 2T \int_0^t dt' \Gamma_{\sigma}(t')
          \left[ e^{2\mu (t-t')} \right]_J       \label{gamas} \\
\Gamma_{\tau}(t) & = & 2\left(\frac{I_1(4t)}{2t}-\frac{e^{-2t}}{2}\frac{I_1(2t)}{t}
                   \right)+2T \int_0^t dt' \Gamma_{\tau}(t')
                   \left[ e^{2\mu (t-t')} \right]_J       \label{gamat} 
\end{eqnarray}
We have solved equations (\ref{gamas}), (\ref{gamat}) and (\ref{overindep})
numerically and the result for the long time limit is shown in Fig.(2).


The final distance decays with temperature more like in the case of opposite initial
conditions than in that of independent staggered ones, at each temperature the
distance is greater than for staggered configurations.
It may be noted that while in the case of staggered configurations both initial
states would lead to non-equilibrium dynamics (zero projection on the maximum
eigenvalue of {\it J}), in this case one replica has a component in the direction $a=2$ while the other has not.

\subsection{Arbitrarily correlated initial conditions}
As a final class of initial conditions we choose a pair that is arbitrarily
close at $t=0$. We considered
$\sigma_{\mu}(0)=1 \forall \mu$, i.e. a uniform initial condition in the firt
replica and for the other we choose
\begin{equation}
\tau_{\mu}(0)= \left\{ 
\begin{array}{r@{\quad if \quad}l}
\kappa &  -2 \le \mu \le 2-\epsilon \\
0      &  2-\epsilon < \mu \le 2 
\end{array} \right. \label{init}
\end{equation}
In this form both initial configurations are equal if $\epsilon=0$ and have a
decreasing correlation as $\epsilon$ grows ($0 \le \epsilon < 4$). The value of
$\kappa$ as a function of $\epsilon$ can be determined exactly from the spherical
constraint. It turns to be $\kappa^2=2\pi/\left[\pi+2\arcsin{(1-\epsilon/2)}+
(1-\epsilon/2)\sqrt{\epsilon(4-\epsilon)}\right]$. Now we have to solve the full
equations (\ref{overlap}) and (\ref{gama}). We note that in this
case the behaviour of $Q(t)$ at $T=0$ is different from the previous cases. 
For correlated initial conditions the distance is not frozen even at zero
temperature. The expressions for the $\Gamma(t)'s$ and $Q(t)$ simplify to:
\begin{eqnarray}
\Gamma_{\sigma}(t)&=&\left[ e^{2\mu t} \right]_J \\
\Gamma_{\tau}(t)&=&\left[ (\sigma_{\mu}(0))^2 e^{2\mu t}\right]_J \\
Q(t)&=&\frac{\left[\sigma_{\mu}(0)\tau_{\mu}(0)e^{2\mu t}\right]_J}
            {\sqrt{\Gamma_{\sigma}(t)\Gamma_{\tau}(t)}}
\end{eqnarray}
It is straightforward to see that
\begin{eqnarray}
Q(t) & = & Q(0)\sqrt{\frac{\Gamma_{\tau}(t)}{\Gamma_{\sigma}(t)}} \nonumber \\
     & = & \frac{1}{\kappa} \sqrt{\frac{\Gamma_{\tau}(t)}{\Gamma_{\sigma}(t)
}} 
\end{eqnarray}
Consequently if $\epsilon \ne 0$, $Q(t)$ evolves in time even at $T=0$. 
An explicit calculation shows that for any
$\epsilon \ne 0, \Gamma_{\tau}(t)$ goes asymptotically to zero
while $\Gamma_{\sigma}(t)$ grows exponentially for long times so that the 
overlap decays exponentially fast and the final distance
is always maximal $D(t)\rightarrow 0.5$.
This is somewhat surprising as it indicates that configurations with arbitrary
initial correlation and deterministic dynamics decorrelate completely 
exponentially fast, something reminiscent of chaotic dynamics. In fact this
behaviour can be understood looking at the geometry of phase space. At zero
temperature the dynamics is deterministic and the asymptotic evolution 
dominated by the maximum eigenvalue. Consequently, a uniform initial condition
 will evolve towards a minimum. The initial condition on the other replica has
zero projection onto the minima and its dynamics will evolve towards the
direction of the maximum eigenvalue onto which it has an initial nonzero
component.
Asymptotically it will reach a fixed point that will correspond, in general, to
a critical point of high order and orthogonal to the minima, then the distance
will be maximal. Conversely, a similar calculation considering two initial 
conditions with
finite projection on the maximum eigenvalue direction shows that the 
configurations asymptotically merge and the distance goes to zero.

For the case of finite temperature we have solved the full equations 
numerically and the result, for initial conditions of the form
(\ref{init}), is identical to that of  Fig.(2) corresponding to
independent uniform initial conditions. So, even at finite temperature, the
distance grows and reaches a finite asymptotic value that depends on the
temperature and is independent of $\epsilon$. 
It can be said that a little initial ``damage'' in a configuration
propagates with time, the phase is ``active''.

$\epsilon$ introduces a time scale for the
decay of the correlations. In general the overlap enters its asymptotic regime in
a time that scales as $t \approx 1/\epsilon$. As $\epsilon$ becomes smaller the
initial conditions have a larger correlation and the times needed for reaching the asymptotic regime grow as $\epsilon^{-1}$.

When both configurations present an initial projection in the direction of the
maximum eigenvalue the asymptotic distance vanishes also in the presence of
thermal noise and a little damage finally dissapears.

\section{Conclusions}

We have studied analytically the spreading of damage in the spherical SK model
with Langevin dynamics. With respect to the general spreading of damage
phenomenology, we have found that in the low temperature phase of the
model the long time distance between the states of two real replicas of the
system depend on the initial conditions on each replica. In this model the
intermediate phase as described in the introduction is absent and the
dynamic transition temperature at which the asymptotic distance begins to be 
non zero coincides with the static one.

The results for zero temperature can be readily interpreted in terms of the
geometry of phase space. The most interesting
result is that two states that are initially arbitrarily close will flow to
a maximal distance if at least one of them has vanishing initial projection 
onto the direction of the maximum eigenvalue. Conversely, if both have an
initial component on the direction of $\mu=2$ the trajectories will finally
merge independently of the initial correlation.

For the other classes of initial conditions considered we found that opposite
or independent ones are fixed points for the dynamics of the distance and are
in this sense less interesting.

At finite temperature the asymptotic distances are simple functions of 
$q_{EA}$ for those
initial conditions that lead asymptotically to equilibrium dynamics, i.e. those
which present an initial condensation in the direction of the maximum 
eigenvalue. In these cases, when the inital configurations are arbitrarily
close to each other they finally merge giving an asymptotic zero distance.

A final comment regards the particular choice for the dynamics. It is known from
the literature on spreading of damage that different dynamics can lead to different
results and so the conclusions can be model dependent. The present 
calculations for Langevin dynamics could be extended to Monte Carlo dynamics 
following the lines of a recent work 
(Bonilla, Padilla, Parisi and Ritort 1996). In that
work it is enphasized the importance of the acceptance rate in the
off-equilibrium dynamics of the system. For asymptotically long times, those
which we are interested in this approach, both dyamics are equivalent and 
one expects that they will give the
same results. The scale of times in which both dynamics are comparable depends on the acceptance rate of the Metropolis algorithm.

The present analysis can be extended to more complicated systems which are 
expected to have a richer phenomenology. The natural extension is to
consider the p-spin spherical model for $p>2$. The study of these models is
in progress.

\vspace{1cm}
{\bf Acknowledgments} I wish to thank D. Dean and J. Kurchan for discussions
and comments.

\newpage
{\Large \bf Figure Captions}\\

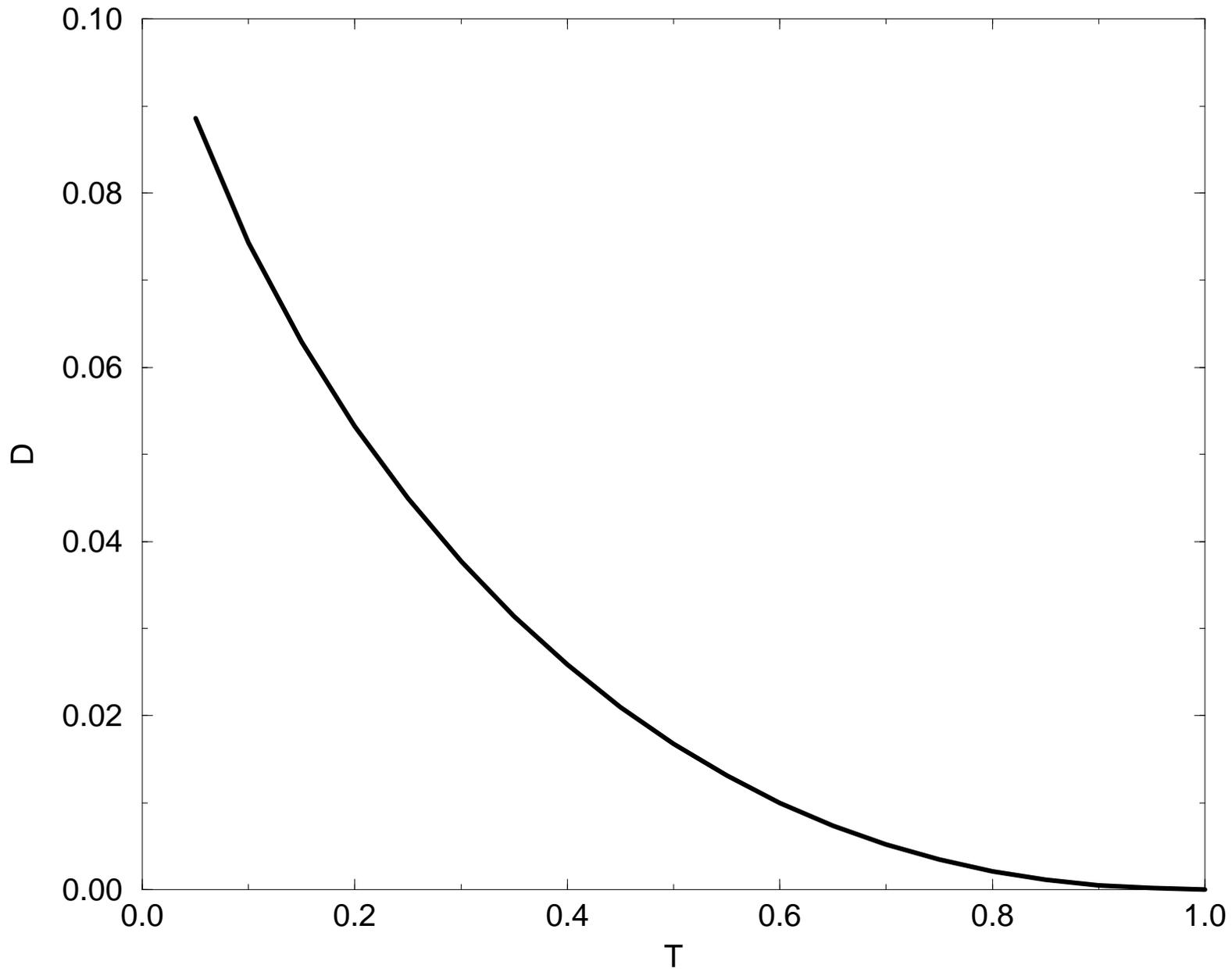
Figure 1: Asymptotic distance vs. temperature for staggered independent initial
conditions with $a=-0.2$ and $b=1.5$ (see text).\\

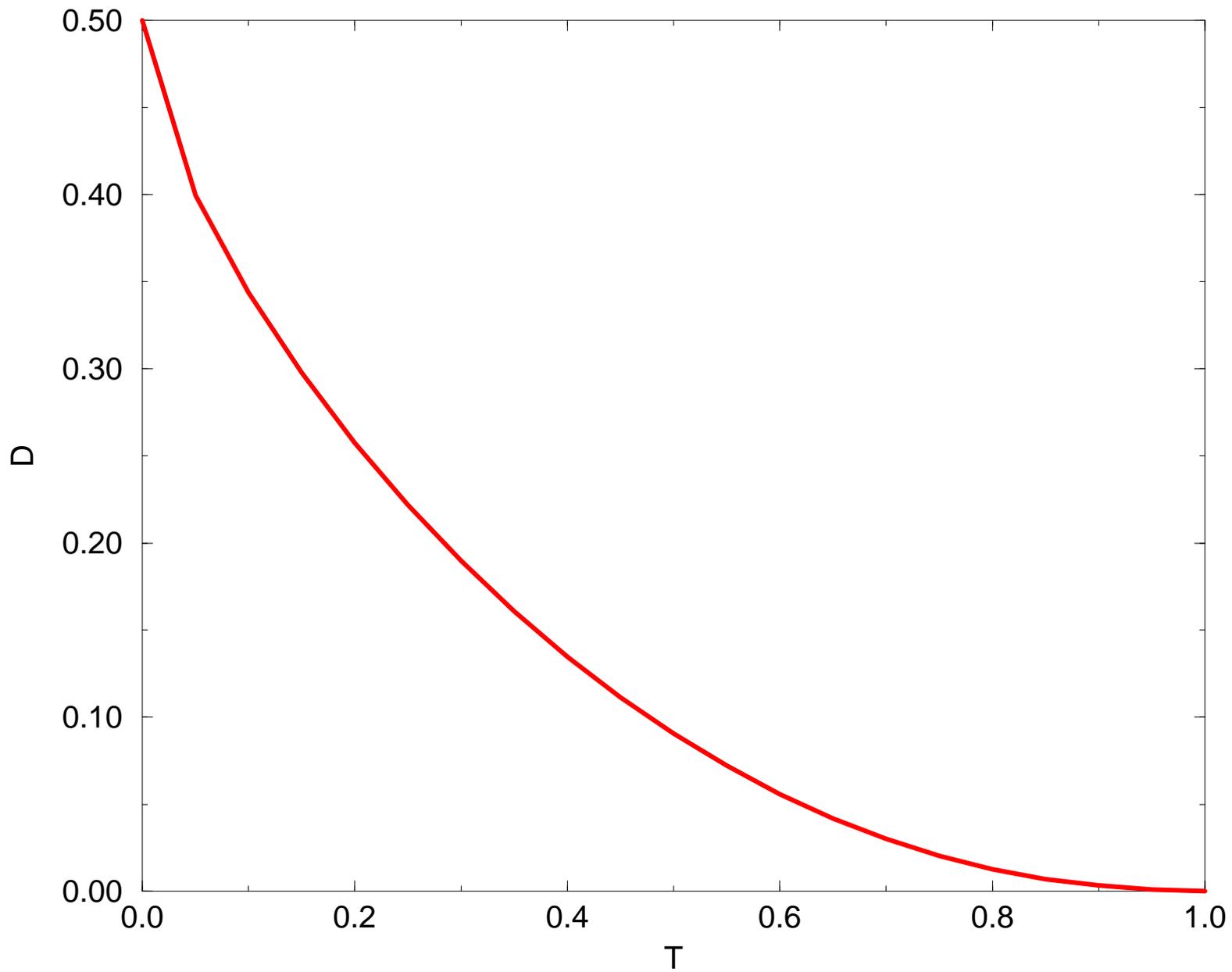
Figure 2: Asymptotic distance vs. temperature for piecewise uniform 
independent initial conditions.

\newpage
{\Large \bf References}

\vspace{2cm}
\flushleft

Bonilla L L, Padilla F G, Parisi G and Ritort F 1996, cond-mat 9602147. \\
Campbell I. A. and de Arcangelis L. 1991 {\it Physica A} {\bf 178}, 29.\\
Coniglio A., de Arcangelis L., Herrmann H.J. and Jan N.
1989 {\it Europh. Lett.} {\bf 8}, 315.\\
Cugliandolo L F and Dean D S 1995 {\it J. Phys. A: Math.Gen.} {\bf 28}, 4213.\\
de Almeida R.M.C., Bernardi L. and Campbell I.A. 1995 {\it J.
Phys. I France} {\bf 5}, 355.\\
Derrida B. 1987 {\it J. Phys. A: Math.Gen.} {\bf 20}, L721.\\
Derrida B. and Weisbuch G 1987 {\it Europhys.Lett.} {\bf 4}, 657.\\
Derrida B. 1989 {\it Phys. Rep.} {\bf 184}, 207.\\
Golinelli O. and Derrida B. 1988 {\it J. Phys. France} {\bf49}, 1633.\\
Hinrichsen H, Weitz J S and Domany E 1996 preprint cond-mat 9611085.\\ 
Kauffman S. A. 1969 {\it J. Theor. Biol.} {\bf 22}, 437.\\
Kosterlitz J M, Thouless D J and Jones R C 1976 {\it Phys. Rev. Lett.} 
{\bf 36}, 1217.\\
Kurchan J and Laloux L 1996 {\it J. Phys. A: Math.Gen.} {\bf 29},1929.\\
Linz P 1985 {\it Analytical and Numerical Methods for Volterra
Equations} (SIAM Studies in Applied Mathematics, Philadelphia).\\
Martins M L, Verona de Resende H F, Tsallis C and
de Magalh\~aes A C N 1991 {\it Phys. Rev. Lett.} {\bf 66}, 2045. \\
Stanley H E, Stauffer D, Kertesz J and Herrmann H J 1987
{\it Phys. Rev. Lett.} {\bf 59}, 2326.\\
Tamarit F A and Curado E M F 1994 {\it J. Phys. A} {\bf 27},671.\\
Vojta T. {\it Damage spreading and dynamical stability of
kinetic Ising models}, cond-mat 9610084;
{\it Chaotic behaviour and damage spreading in the Glauber Ising model -
a master equation approach}, cond-mat 9611232.\\

\end{document}